\documentclass[showpacs,aps]{revtex4}
\usepackage{graphicx}
\begin{document}

\baselineskip 14pt

\title{On topological charge carried by nexuses and center vortices}
\author{John M. Cornwall*}
\affiliation{ Department of Physics and Astronomy, University of California, Los Angeles, Los Angeles, California 90095} 
\begin{abstract}
\pacs{PACS number(s):  11.15 Tk, 12.38.-t \hfill UCLA/01/TEP/35}

In this paper we  further explore the question of topological charge in the center vortex-nexus picture of gauge theories.  Generally, this charge is locally fractionalized in units of $1/N$ for gauge group $SU(N)$, but globally quantized in integral units. We show explicitly that in d=4 global topological charge is a linkage number of the closed two-surface of a center vortex with a nexus world line, and relate this linkage to the Hopf fibration, with homotopy $\Pi_3(S^3)\simeq Z$; this homotopy insures integrality of the global topological charge.    We show that a standard nexus form used earlier, when linked to a center vortex, gives rise naturally to a homotopy $\Pi_2(S^2)\simeq Z$, a homotopy usually associated with 't Hooft-Polyakov monopoles and similar objects which exist by virtue of the presence of an adjoint scalar field which gives rise to spontaneous symmetry breaking.  We show that certain integrals related to monopole or topological charge in gauge theories with adjoint scalars also appear in the center vortex-nexus picture, but with a different physical interpretation.  We  find a new type of nexus  which can carry topological charge by linking to vortices or carry d=3 Chern-Simons number without center vortices present; the Chern-Simons number is connected with twisting and writhing of field lines, as the author had suggested earlier.  In general, no topological charge in d=4 arises from these specific static configurations, since the charge is the difference of two (equal) Chern-Simons number, but it can arise through dynamic reconnection processes.    We complete earlier vortex-nexus work to show explicitly how to express globally-integral topological charge as composed of essentially independent units of charge $1/N$.

\end{abstract}
\thanks{Email address:  Cornwall@physics.ucla.edu }
\maketitle

\section{Introduction}

In this paper, which investigates the properties of topological charge in the center vortex-nexus picture of QCD-like gauge theories, the reader will recognize a number of familiar formulas, but possibly not the context in which they arise.  These familiar formulas are well-known in what we will call the traditional view \cite{cj,gpy,ct1,ct2,ct3,dgo,vb} of monopoles and topological charge in gauge theories, in which some sort of adjoint scalar field is invoked (such as a gauge-fixing field,  the time component of the gauge potential, or a Polyakov line), leading then to BPS or 't Hooft/Polyakov monopoles.  These and similar monopoles are often invoked  as integrally-charged instanton constituents \cite{cg,ht,bs,bot,bcnot,sm,oj}.  Sometimes discussed in the traditional viewpoint are fractional topological charges.  For example, in periodic (finite-temperature) space there are  calorons \cite{vb}, whose topological charge unit is $1/N$ for $SU(N)$, as instanton constituents.  

The trouble with using adjoint scalars as a tool for investigating QCD-like gauge theories is that
if the scalar field is elementary one is not dealing with QCD; if it is not, one is trying to explain general features of QCD from a gauge- or frame-dependent point of view. 
In the center vortex-nexus view \cite{ct,co95,agg,c98,c99,er00,co00,cpvz,bvz} of gauge theories, explored in the present paper, there are no adjoint scalar fields and gauge-fixing is unnecessary. (The name nexus is not always used; Refs \cite{cpvz,bvz} call them center monopoles.)  Topological charge is composite and expressed by the linkages of nexus world lines and center vortex surfaces \cite{co95,er00,co00}.  We show that the composite nature of topological charge is in some sense equivalent to having an adjoint scalar, and that some of the same mathematical formulas of the traditional monopole picture emerge in identical or near-identical form in the nexus-vortex picture, but with different physical and mathematical interpretation. For example,  the same surface integral which  expresses topological charge as Abelian monopole charge also arises with vortices and nexuses, but the surface has physical meaning as a vortex surface.  Other topological formulas familiar in the traditional viewpoint arise, 
even in the absence of symmetry-breaking scalars. One may explain global integrality of topological charge through the homotopy $\Pi_2(S^2)\simeq Z$, which arises for vortices and nexuses as well as for  traditional monopoles.

Although global topological charge is certainly integral (in a compactified space), in the center vortex-nexus picture it consists of a set of localized fractional charges which are topologically confined.  So there is considerable structure in a unit of topological charge.   
For almost as long as there have been instantons, there has been interest from the traditional point of view as well as to whether there is deeper structure to the topological charge carried by instantons at the classical level.  The possibility of fractional topological charge in two-dimensional $CP^N$ models was discussed long ago\cite{ffs}, and in d=4 gauge theories this has been discussed by numerous authors (a sample of such works prior to recent center vortex-nexus activity is  \cite{w79,th81,ct,ehn}).  There is also the suggestion that, in various gauges involving adjoint scalars, monopoles such as BPS or 't Hooft-Polyakov monopoles are instanton constituents \cite{cg,ht,bs,bot,bcnot,sm,oj}.  These monopoles typically have integral topological charge.  The suggestion that monopoles carry topological charge goes back even further, for example, Refs. \cite{cj,gpy,ct1,ct2,ct3,dgo}. 

In general, the work of the traditional viewpoint is based on a combination of certain ingredients.  These ingredients include an adjoint scalar field, which may be an elementary Higgs-like field; the $A_0$ component of the gauge potential; a Polyakov line; or an auxiliary field introduced to implement certain Abelian gauges of the type first discussed by 't Hooft, including the maximal 
Abelian gauge or the Laplacian Abelian gauge.  In some works the discussion is restricted \cite{cj,gpy} to a space-time or Euclidean space with one or more compact dimensions, such as $R^3\times S^1$, and the instanton is a periodic instanton such as a caloron \cite{vb}; in other works, the picture of instanton constituents depends essentially on a choice of gauge.  Since the adjoint scalar (generically) breaks an $SU(N)$ gauge symmetry to $SU(N)/U(1)^{N-1}$, monopole-like homotopies such as $\Pi_2(SU(N)/U(1)^{N-1})=(Z)^{N-1}$ are relevant, and the integral winding numbers of these homotopies are identified, typically, with the magnetic charges of BPS monopoles which exist because of the adjoint scalar field.  (In the case of maximal Abelian gauges 't Hooft pointed out that degeneracies of eigenvalues of certain local gauge conditions led to Abelian monopoles; these are essentially equivalent to the BPS monopoles.)  

While all this work with adjoint scalars or periodic dimensions is certainly intriguing, it has some drawbacks.  First, one would like to understand what is going on in plain old QCD, that is, in space-times with no periodic dimensions and without the crutch of adjoint scalars (even non-elementary adjoint fields such as $A_0$, because  Lorentz or Euclidean invariance forbids a global expectation value for $A_0$) and no symmetry breaking. Correspondingly it would appear that there is no $\Pi_2$ homotopy of the usual sort.  How, then, can one interpret the results discussed above, which make use of ingredients not possessed by QCD?  Or is this even possible?  Does the presence of scalar fields lead to results which must differ essentially from those of QCD?  Long ago, Witten \cite{wit79}  expressed exactly this view: There is a fundamental difference between gauge theories having scalar fields and symmetry breaking and those without scalars.  The former have integrally-charged instantons for reasons having to do with single-valuedness of the scalar fields, but there is no reason for the latter to have integral lumps of topological charge, in the presence of quantum gauge-field fluctuations.  Witten was drawn to such a conclusion in part by wondering how the $\eta'$ mass arose, which soon led him to the Witten-Veneziano sum rule \cite{w79,v79}.  This relation between the $\eta '$ mass and the topological sucsceptibility strongly suggests that in $SU(N)$ the unit of topological charge is $1/N$, even though the QCD vacuum has period $2\pi$ in the $\theta$ angle and even though the total topological charge in a compact space such as $S^4$ is necessarily integral. Our view is that in QCD-like gauge theories topological charge may not be carried by instantons, but it is still possible to understand integral topological charge as a sort of topological confinement of fractional constituents.

Clearly the major differences between the two viewpoints summarized above is that there is no need for any adjoint scalar field or effective symmetry-breaking in the center vortex-nexus picture, and no need for tying the results obtained to any particular gauge, such as a maximal Abelian gauge.  These gauges in effect break the gauge symmetry, as noted above, leading to homotopies such as   $\Pi_2(SU(N)/U(1)^{N-1})=(Z)^{N-1}$.     
As the appearance of a $\Pi_2$ homotopy suggests, the magnetic charge of a monopole is measured by an integral over a closed 2-surface.  There is nothing special about this, however.  The long exact sequence of homotopy says that for a compact simply-connected Lie group $G$ with subgroup $H$,
\begin{equation}
\label{exseq}
\Pi_n(G/H)\simeq \Pi_{n-1}(H_0)
\end{equation}
where $H_0$ is the component of $H$ connected to the identity in $H$.  So when a gauge symmetry is spontaneously broken a $\Pi_2$ homotopy ($n=2$) implies a $\Pi_1$ homotopy, which can be identified with loops around Dirac strings.  

In the vortex-nexus picture we will show that one can identify topological charge with  a sort of magnetic charge of a nexus, measured either as a surface integral of the type associated with $\Pi_2$ homotopy or with loops around strings.  The surface is not just the mathematical surface at infinity, but is instead a physical surface, the two-surface of a center vortex which is linked to a nexus world line.  (It is, in effect, at infinity because entropy causes vortex surfaces and nexus world lines to extend to great distances compared to the inverse mass scale of QCD.)  The strings are fat, have no short-distance singularities, and there is no symmetry breaking. 
We show that certain maps and homotopies can be associated with topological charge in the center vortex-nexus viewpoint.  Of course, all the homotopies are  the same, that is, the group of integers $Z$.  The discussion of homotopies begins by approximating  the fat strings of physical vortices and nexuses by Dirac strings which appear in the pure-gauge limit. We then replace these Dirac strings by smooth Abelian maps with the same magnetic flux or topological charge, using a variant of old ideas about monopole charge \cite{afg} and
the work of \cite{oj,jp}, which shows how to map  non-Abelian  potentials and fields onto Abelian potentials and fields with the same topological charge (see also the closely-related Ref. \cite{km}).  This discussion of homotopy bypasses certain tedious mathematical questions by invoking Dirac strings instead of multiply-connected spaces.
  
  The usual monopole homotopy relation  $\Pi_2(S^2)\simeq Z$ arises, and is  associated with a typical monopole map $S^2\rightarrow S^2$ of a unit vector field associated with a nexus onto a closed two-surface associated with a plain center vortex.  This homotopy is, of course, consistent with the  requirement of integral topological charge in a compact d=4 space.  The novel point here is that there is no adjoint scalar field, which usually carries the monopole charge.  The fractionalization of topological charge constituents is expressed through a $\Pi_1$ homotopy of center vortices, following from the fact that the adjoint representation of the gauge group is $SU(N)/Z_N$.  This homotopy, $\Pi_1(SU(N)/Z_N)\simeq Z_N$, expresses the content of confinement by center vortices.   For confinement, it refers to wrapping a Wilson loop around (in d=3; or linking to the surface of, in d=4) a center vortex, but for topological charge, it refers to wrapping (part of) a vortex surface around a nexus Dirac string.

We will also show explictly how the usual d=4 topological charge integral over $G\tilde{G}$ becomes a linking number of a center-vortex surface with a nexus world line, weighted with trace factors which are integers.  This link number can be given an expression in d=3, where we argue that this is the linkage of  the Hopf fibration of $S^3$ into a base space $S^2$ and a fiber $S^1$:  $\Pi_3(S^2)\simeq Z$.  A standard theorem says that this homotopy is isomorphic to the usual topological-charge homotopy $\Pi_3(S^3)\simeq Z$.  Many authors have discussed physical applications of this homotopy \cite{ct1,ct2,ct3,dgo,vb,kvb,pvb,bhvw,oj,km,jp} in the traditional viewpoint.

We exhibit a new form of nexus, which carries topological charge or Chern-Simons number in several different ways.  First, it links to a vortex in the same way as the original nexus.  Second, as a static configuration in d=3, it carries integral Chern-Simons number with no reference to vortices. As a purely static configuration it cannot yield d=4 topological charge, which is the difference of two (equal, in this casse) Chern-Simons numbers.  Third, it can be extended to a dynamic configuration which does carry d=4 topological charge, but without linkage to vortices; topological charge arises through reconnection processes which change link, or its equivalents, twist and writhe.   These processes have been outlined in earlier works \cite{cy,co96,co98}, but not connected to configurations such as this new nexus.

Finally, we give an explicit discussion of the decomposition (mentioned but not carried out in \cite{co00}) of a unit of topological charge into $N$ units, each of charge $1/N$.

In Section \ref{nexus} we give a brief review of the nexus-center-vortex picture of confinement and of fractional topological charge.  Section \ref{interpret} gives the interpretation of the usual definition of topological charge as the integral over $G\tilde{G}$ in terms of an integral of the magnetic flux of a monopole-like configuration (the nexus) over a closed 2-surface and compares the result to previous investigations of BPS monopoles.  Section \ref{homo}  remarks on various maps and homotopies of the center vortex-nexus interaction, including the Hopf fibration and its interpretation as a link number.  Section \ref{anewnexus} outlines the properties of the new nexus, which  can link to vortices or carry Chern-Simons number without linkage to a separate vortex.  Section \ref{frac} show how to divide a charge-one instanton into  $N$ fractional charges.    Section \ref{conc} gives a summary.

\section{A review of nexuses and center vortices}
\label{nexus}

Nexuses and center vortices can be defined on the lattice or in the continuum, and we restrict ourselves to the continuum approach.  The idea is that infrared instability of QCD (in d=3,4) forces the Schwinger-Dyson equations for the pinch-technique (gauge-invariant) gluon propagator to have only massive solutions, describing a constituent gluon mass which scales with the RG mass $\Lambda_{RG}$.  There is an infra-red effective action describing this phenomenon, which is just the usual action for the gauge field plus a gauged non-linear sigma-model action.  Although the latter action possesses the necessary quality of gauge invariance, it is not renormalizable as a fundamental interaction.  Nonetheless, the total action serves as an accurate description of the infrared quantum properties of QCD.  Renormalizability of the underlying theory is assured because the Schwinger-Dyson equations have massive gluons with a mass which vanishes at large momentum like $q^{-2}$ (modulo logarithms).

We will use anti-Hermitean gauge potentials and fields, with the gauge potential matrix described in terms of the canonical potentials and coupling $g$ via:
\begin{equation}
\label{def}
A_{\mu}(x)=\sum \frac{\lambda_J}{2ig}A_{\mu}^J(x),
\end{equation}
and similarly for the gauge field matrix $G_{\mu\nu}(x)$. The generators are normalized conventionally:
\begin{equation}
\label{conv}
Tr\lambda_J\lambda_K=2\delta_{JK}.
\end{equation}
We also introduce a unitary matrix $U(x)$ in the fundamental representation of the gauge group (always taken to be $SU(N)$ in this paper).  Under a gauge transformation $V$, the gauge potential and $U$ transform as:
\begin{equation}
\label{gauge}
A_{\mu}\rightarrow VA_{\mu}V^{-1}+V\partial_{\mu}V^{-1};\;U\rightarrow VU.
\end{equation}
The Euclidean infrared-effective action, containing a constituent gluon mass $m$,
\begin{equation}
\label{action}
S=-\int d^4x[\frac{1}{2g^2}TrG_{\mu\nu}^2+\frac{m^2}{g^2}Tr(D_{\mu}UU^{-1})^2]
\end{equation}
is gauge-invariant.  In spite of appearances, the matrix $U$ is not an independent degree of freedom, since its equation of motion, found by varying the action, is just the identity
\begin{equation}
\label{identity}
[D_{\mu},[D_{\nu},G_{\mu\nu}]\equiv 0.
\end{equation}

\subsection{Center vortices}

This action has numerous static soliton solutions, among them center vortices and nexuses.  The reason for the existence of many solutions for the effective action, while there are none for the classical action, is elementary.  If one chooses any decomposition of the gauge potential in terms of dimension-one functions of the type $f(\lambda \vec{x})/r$, where $\lambda$ is a variational mass scale, and calculates the static (d=3) action, the classical term scales like $\lambda$ and the mass term scales like $m^2/\lambda$.  The sum of the two terms in the action always has a minimum with $\lambda\sim m$.  

Center vortices are expressed in terms of a set of matrices $\{Q_a\}$, $a=1\dots N$ (described below) and a closed 2-surface.  We will only need the explicit form for a center vortex whose holonomy is the generating element of the center of the group $Z_N$, which is $\exp (2\pi i /N)$.  This form is:
\begin{equation}
\label{cv}
A_{\mu}(x)=\left( \frac{2\pi Q_a}{i}\right) \epsilon_{\mu\nu\alpha\beta}\partial_{\nu}\oint \frac{1}{2}d\sigma_{\alpha\beta}[\Delta_m (x-z(\sigma ))-\Delta_0(x-z(\sigma ))].
\end{equation}
Here $\Delta_{m,0}$ is the free propagator for mass $m,0$.  Note that the vortex is not singular on the surface ($x=z(\sigma )$), where the short-distance singularities of the two propagators cancel out.  The matrices $\{Q_a\}$ are diagonal, of the form:
\begin{equation}
\label{qmatrix}
Q_a=\rm{diag}(\frac{1}{N},\frac{1}{N},\dots -1+\frac{1}{N},\frac{1}{N},\dots )
\end{equation}
with the -1 in the $a^{th}$ position.  These matrices have the property that
\begin{equation}
\label{center}
e^{(2\pi iQ_a)}=e^{(2\pi i/N)}
\end{equation}
for any $a$.  The $Q$ matrices are traceless, and the sum of all $N$ of them gives zero. The trace of the product of two $Q$ matrices will be needed; it is:
\begin{equation}
\label{trace}
TrQ_aQ_b=\delta_{ab}-\frac{1}{N}.
\end{equation}

 An important special case of this center vortex is the static solution, with the 2-surface $S$ chosen as the $zt$ plane:
\begin{equation}
\label{static}
\vec{A}(\vec{x})=\frac{\hat{\phi}Q_a}{i}[\frac{1}{\rho}-mK_1(m\rho )],
\end{equation}
where $K_1(m\rho )$ is the Hankel function of imaginary argument.
Vortices corresponding to the element $\exp (2\pi i J/N)$ can be found by replacing $Q_a$ in equation (\ref{static}) by a sum of $J$ different $Q$-matrices.  The fundamental-representation Wilson-loop holonomy $\exp (\oint dx_{\mu}A_{\mu})$ is this $J^{th}$ element of the center raised to a power which is the Gauss linking number of the vortex surface and the Wilson loop; this gives rise to confinement. 

\subsection{Nexuses}

Roughly speaking, a nexus is a closed string lying on a closed center-vortex surface which is the boundary between regions on the vortex surface where the gauge field strength has different orientations, yet this string is indetectable in the Wilson-loop holonomy.   This string can be envisaged as the world line of a nexus which is essentially a point-like static lump; that is, one can think of the nexus as a monopole-like soliton, with its magnetic field strength lying along two tubes of finite thickness instead of spreading out isotropically, as in the 't Hooft-Polyakov monopole.  These tubes are the static projection of the center vortex surface in which the nexus is embedded. For obvious reasons of continuity, if a vortex surface has a nexus it must also have an anti-nexus where the field-strength orientations change to their original values; this closed tube of flux decorated with a nexus and and anti-nexus is the three-dimensional projection of the 2-surface of a center vortex with a nexus and an anti-nexus world line drawn on it.  Since the holonomy remains unchanged as one crosses these lines, if the vortex corresponds to the $J=1$ element of the center then the only allowed changes in field strength are that the $Q_a$ appearing in one region of the surface can be replaced by another, call it $Q_b$, with $a\neq b$ (and similarly for higher values of $J$) in the region reached by crossing the nexus.  Note that for $SU(2)$ this amounts to changing the sign of the field strengths as the nexus is crossed.  Of course, this must happen smoothly and without singularities, so the nexus is not a mathematical boundary, but a string of thickness $\sim m^{-1}$.

While the center vortex can be completely imbedded in an Abelian subgroup, the center vortex is genuinely non-Abelian.  There is no known  explicit solution for a nexus, but it is easy to describe the nexus accurately enough for our purposes. All nexuses can be composed from a fundamental nexus of $SU(2)$.  We describe the static version (analogous to the static center vortex of equation (\ref{static})), in terms of several trial wave functions.  One possible set (not unique) of nexus trial wave functions with flux tubes oriented along the $z$ axis (as for the center vortex of equation (\ref{static})) is:
\begin{equation}
\label{trial}
A_j(\vec{x})=\frac{\epsilon_{jkl}\tau_k\hat{x}_l}{2ir}(F+G\cos \phi -1)+\left( \frac{\tau_j-\hat{x}_j\tau \cdot \hat{x}}{2ir}\right) G\sin \phi+\frac{\hat{\phi}_j\tau \cdot \hat{x}}{2i}(\frac{1}{\rho}-B).
\end{equation}
Here the $\vec{\tau}$ are the Pauli matrices and $F,G$ are dimensionless functions of $\rho ,z$ obeying the boundary conditions
\begin{eqnarray}
\label{bcond}
r\rightarrow \infty :\;\;G\rightarrow 1,\;F\rightarrow 0 \\ \nonumber
r\rightarrow 0:\;\;G\rightarrow 0,\;F\rightarrow 1.
\end{eqnarray}
As for the function $B(\rho ,z)$, it approaches $1/\rho$ for small $\rho$, but has special behavior at infinity.  It vanishes at infinity along any ray of constant $\theta\neq 0,\pi$.  However, it does not vanish as $z\rightarrow \infty$ at fixed $\rho$.  Instead, for large enough $z$, $B$ approaches $mK_1(m\rho )$ as in the center vortex solution (\ref{static}).  The associated anti-nexus is not explicitly described in the trial wave function; it lives at infinite distance.   

Note that the kinematics chosen for the nexus combine some of the standard forms for spherically-symmetric solutions with a form rather like that of a center vortex, except for the group-theoretic factor.  The energy of the trial wave function of equation (\ref{trial}) consists of a vortex-like part coming from the last term on the right in (\ref{trial}), which diverges trivially because of the infinite length of the vortex, plus a part centered on the origin which adds an amount of order $m/g^2$ to the energy.  

The essential ingredients of a nexus-vortex combination can be deduced from the behavior of the trial wave function near infinity, where this wave function approaches that of the singular gauge transformation $V$:
\begin{equation}
\label{veq}
r\rightarrow \infty:\;\;A_j(\vec{x})\rightarrow V\partial_jV^{-1},\;V=e^{(i\phi \tau\cdot\hat{x}/2)}.
\end{equation}
This is the same as the $m\rightarrow \infty$ limit of the wave function, analogous to the  $m\rightarrow \infty$ limit of the pure center vortex given in equation (\ref{minf}) below.  The  holonomy of a large Wilson loop far from the origin and encircling the $z$ axis once (or an odd number of times) is just
\begin{equation}
\label{holo}
V(\phi =2\pi )V^{-1}(\phi =0)=-1.
\end{equation}
This is, of course, the holonomy of the center vortex in which the nexus is embedded.

The gauge transformation $V$ of equation (\ref{veq}) does not approach $\pm I$ at infinite distance, which one ought to require in order to compactify $R^3$ to $S^3$ which insures integral topological charge.  Of course, there is a compactified form of (\ref{veq}), in which the nexus and the anti-nexus live on a compact closed vortex.  But we will not have need of any explicit expression for this case.  

There are infinitely-many other pure-gauge nexus forms having this holonomy property.  They differ from (\ref{veq}) above in that the unit vector $\hat{x}$ is replaced by any other unit vector $\hat{e}$ having the property that at $x,y=0$ this vector points along $\hat{z}$ for $z> 0$ and along $-\hat{z}$ for $z<0$.  We will consider other forms for $\hat{e}$ in discussing homotopies of topological charge.

The gauge transformation $V$ has Dirac strings which are the $m=\infty$ limit of the smooth flux tubes of the full nexus (or center vortex).  The Dirac-string field strengths are easily calculated:
\begin{equation}
\label{dirac}
\vec{G}=-\left( \frac{\tau_3}{2i}\right) 2\pi \hat{z}\epsilon (z)\delta (x) \delta (y).
\end{equation}
There is a string going left and one going right, and unlike a center vortex, the field strengths point in opposite directions along the two string.  This is because the $\tau \cdot \hat{x}$ appearing in the gauge transformation $V$ become $\tau_3\epsilon (z)$ when multiplied by $\delta (x)\delta (y)$.
 
The generalization of this to arbitrary $N$ is straightforward, using formulas previously given \cite{co00}, and will be discussed in Section \ref{frac}.  It simply depends on recognizing that transformations changing $Q_a$ into $Q_b$ can always be embedded in an $SU(2)$ lying in the 2$\times$2 $a,b$ subspace.

Although strictly speaking a nexus is the localized region on a nexus-vortex combination where the field lines change direction in the gauge group, we will from now on use the word nexus to refer to the nexus itself plus the vortex sheets attached to it.

\section{Topological charge as an integral of magnetic flux}
\label{interpret}

\subsection{Topological charge as a modified intersection number}

As explained in \cite{co00,er00}, the usual topological charge $Q$:
\begin{equation}
\label{q}
Q=\frac{-1}{16\pi^2}\int d^4x Tr G_{\mu\nu}\tilde{G}_{\mu\nu}
\end{equation}
can be expressed as a weighted intersection number of the intersections of the closed 2-surfaces forming center vortices and nexuses.  The weights come from traces over group-theoretic factors such as the $\{Q_a\}$ of Section \ref{nexus}.  Since by entropic arguments the surfaces have scales very large compared to $m^{-1}$ it is sufficient to express $Q$ in terms of the  vortices and nexuses with $m=\infty$, in which case a center vortex corresponding to the fundamental center element $\exp (2\pi i/N)$ becomes:
\begin{equation}
\label{minf}
\tilde{G}(x)_{\mu\nu}=\left(\frac{2\pi Q_a}{i}\right)\int d\sigma_{\mu\nu}\delta (x-z(\sigma ))
\end{equation}
where any value of $a$ can be used.  (The general expression in the $m\rightarrow \infty$ limit for a nexus is somewhat more elaborate; see \cite{co00}.  We will only need certain special cases of the general explicit expression in the present paper.)  One readily sees that $Q$ is expressed as an intersection number of surface elements:
\begin{equation}
\label{qno}
Q=\sum_{A,B}2Tr(Q_AQ_B)I(A,B)
\end{equation}
where the label $A,B$ of surface elements includes the geometric description of the surface as well as the choice of $Q_a$ for the surfaces, and the sum counts distinct intersections only once.  The intersection number $I(A,B)$ is:
\begin{equation}
\label{xno}
I(A,B)=\epsilon_{\mu\nu\alpha\beta}\int \frac{1}{2}d\sigma_{\mu\nu}\frac{1}{2}d\sigma_{\alpha\beta}\delta (x(A)-x(B)).
\end{equation}

Since $Tr(Q_AQ_B)=\delta_{AB}-1/N$ it might appear that the total topological charge integrated over all $d^4x$ could be fractional.  Actually, this is not so if only vortices without nexus world lines appear in $Q$.  This is because the surfaces to be integrated over are {\em closed} and for every intersection in (\ref{xno}) occurring with a positive sign there is another one occurring with negative sign, so everything cancels to zero.  (A simple demonstration of this will be given below.) But the situation changes when nexuses are involved, since if the nexus world line is {\em linked} to a plain center vortex the traces do not cancel; instead they automatically reduce to integral values \cite{co00,er00}.

\subsection{Topological charge as a surface integral}

The interpretation of topological charge as the magnetic flux or charge of a 't Hooft-Polyakov-like monopole is quite old (see, {\it e.g.}, Refs. \cite{cj,gpy,ct1,dgo}).  These authors showed that, given one simple monopole and an Abelian projection $P$ which in certain cases can be identified with the unit vector corresponding to the VEV of an adjoint scalar field, the topological charge had an expression as an integral over the two-surface at spatial infinity. (We need not discuss here the particular boundary conditions and scalar fields which enter the discussion for each of the cited works.)  This integral is: 
\begin{equation}
\label{cjeq}
Q=\frac{1}{4\pi i}\int d\vec{S}\cdot Tr(P\vec{B})
\end{equation} 
where $\vec{B}$ is the magnetic field of the monopole.  The 2-surface of integration was the surface at $r=\infty$.

In view of the absence of adjoint scalar fields  (therefore of 't Hooft-Polyakov monopoles) it might seem unlikely that the integral (\ref{cjeq}) could apply to QCD, since not only is there no obvious candidate for a $\Pi_2$ homotopy, but also the magnetic charge of a monopole can be expressed entirely in terms of the adjoint scalar field \cite{afg}.  We will now show that this integral does occur, provided that the 2-surface of integration is interpreted as the surface of a plain center vortex.

In the defining equation (\ref{q}) for topological charge, consider the contribution of a nexus $A$ linked to a plain center vortex $B$.  These contribute a term
\begin{equation}
\label{ab}
\frac{-1}{8\pi^2}\int d^4xTr\tilde{G}_{\mu\nu}^{(B)}G_{\mu\nu}^{(A)}
\end{equation}
to the topological charge.  Replace $\tilde{G}^{(B)}$ by its infinite-mass limit in equation (\ref{minf}) to find:
\begin{equation}
\label{q3}
Q=\frac{1}{4\pi i} \int d\sigma_{\mu\nu}Tr(Q_bG_{\mu\nu}^{(A)}).
\end{equation}
Here we associate the matrix factor $Q_b$ with the center vortex $B$.  

Without affecting any topological conclusions, we can always choose a vortex and nexus  such that the center vortex surface lies in a 3-space $(x_1,x_2,x_3)$ and the nexus surface is the product of a closed line in this 3-space times the remaining coordinate $x_4$. That is, the nexus is displayed as a static soliton of the type in equation (\ref{trial}).    Then, using conventional three-space notation, the above integral for the topological charge looks like the traditional expression for monopole charge:
\begin{equation}
\label{charge}
Q=\frac{1}{4\pi i}\int d\vec{S}\cdot 2Tr(Q_b\vec{B}^{(A)}).
\end{equation}  

\begin{figure}[ht]
\includegraphics{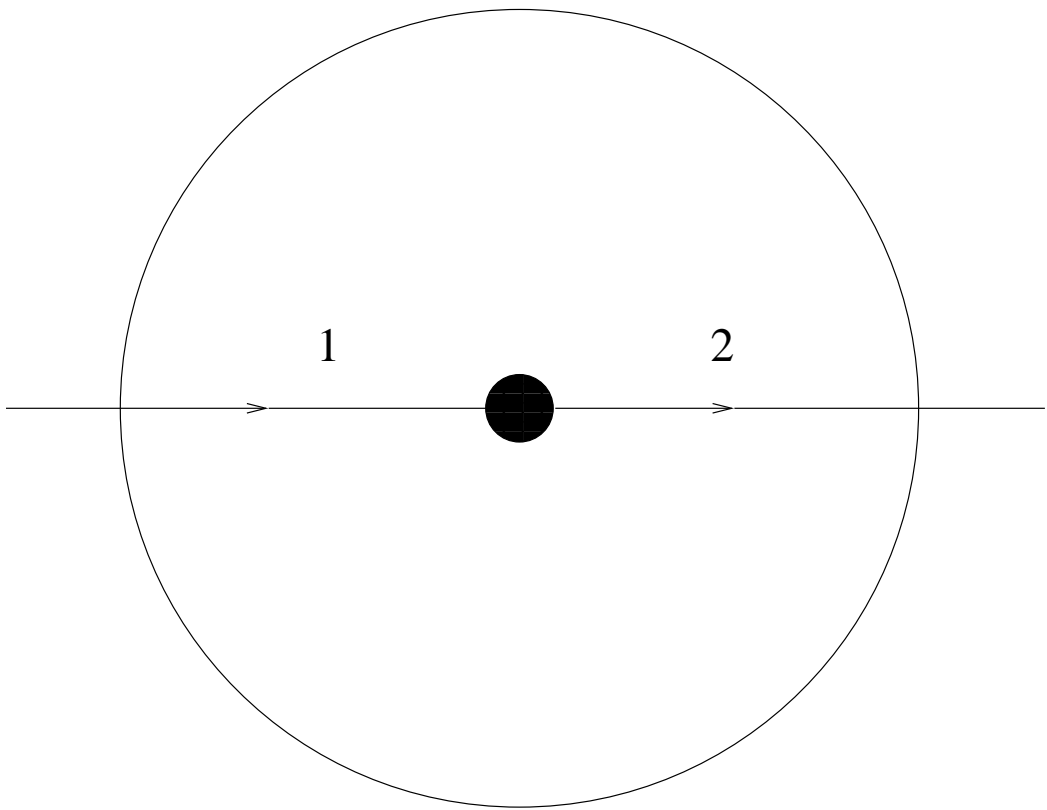}
\vskip.2cm
\caption[]{
\label{nexusfig1}
\small The outer circle represents the plain vortex surface with flux described by $Q_2$.  The inner black dot represents a nexus, and the lines represent the associated flux tubes, with fluxes described by $Q_{1,2}$.}
\end{figure}

In view of equation (\ref{q2}), the flux label $b$ must be the same as one of the two flux labels on the nexus or else the topological charge is zero.  For $SU(2)$ the only nexus has labels (1,2) and we will choose $b=2$ to be definite.  The configuration of vortex intersecting with a nexus and its flux tubes is shown schematically  in Fig. \ref{nexusfig1}. (The circle in the figure is not to be taken too literally; surfaces of different genus may occur.) In $SU(2)$ since $Q_1=-Q_2=\tau_3/2$, the trace factor in equation (\ref{charge}) including the factor of 2 can be written as $Tr\tau_3(\tau_3/2)$ where the first $\tau_3$ is the projector $P$ of equation (\ref{cjeq}).    Roughly speaking, this factor of $\tau_3$ in equation (\ref{charge}) substitutes for the Higgs field in equation (\ref{cjeq}). Using the nexus field of equation (\ref{dirac}) yields unit topological charge:
\begin{equation}
\label{chargeg}
Q=\frac{1}{4\pi }\int d^2\vec{S}\cdot Tr(\tau_3\vec{G}).
\end{equation}
 
\begin{figure}[ht]
\includegraphics{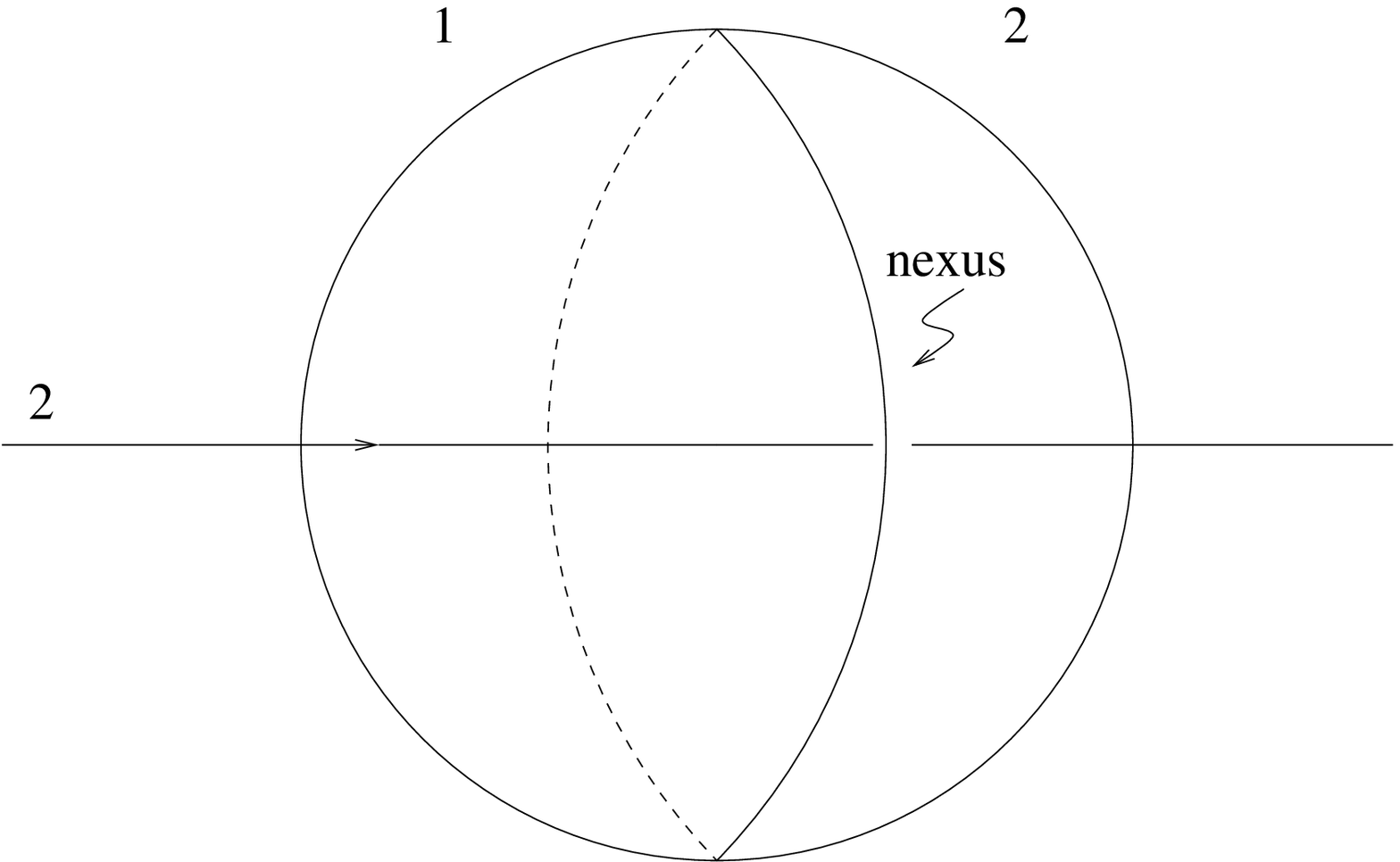}
\vskip.2cm
\caption[]{
\label{nexusfig1a}
\small An alternate view of topological charge formed by the intersection of a plain vortex and one with a nexus world-line on it.  As before, the fluxes are labeled by the numbers 1,2.}
\end{figure}

Of course, we could equally well choose to display the center vortex with the nexus on it as a closed surface, with the nexus world-line linked to a plain center vortex (which appears as a simple line in projection).  This is shown schematically in Fig. \ref{nexusfig1a}.   In this figure we do not display the anti-nexus which necessarily accompanies the nexus, and so the figure is just a schematic.

As usual, the fact that the total topological charge is integral can be ascribed to the operation of certain homotopies and maps.  We discuss these issues next.

\section{Vortex-nexus interactions and homotopy}
\label{homo}

We discuss in this section two maps and their homotopies which are relevant to the integrality of the total topological charge, aside from the conventional $\Pi_3(S^3)\simeq Z$ homotopy of the standard $G\tilde{G}$ charge integral, or equivalently a difference of Chern-Simons integrals. One map is the standard monopole map $S^2\rightarrow S^2$, and the other is the Hopf fibration $S^3\rightarrow S^2$.  The  homotopy $\Pi_3(S^2)$ of the latter map is the same as  $\Pi_3(S^3)\simeq Z$ by the exact sequence for a fiber bundle $E$ with base space $B$ and fiber $F$ (a special case of which is used in equation (\ref{exseq}):
\begin{equation}
\label{seq}
\rightarrow \Pi_{n+1}(B)\rightarrow \Pi_n(F)\rightarrow \Pi_n(E) \rightarrow \Pi_n(B)\rightarrow \Pi_{n-1}(F)\rightarrow \cdots
\end{equation}
plus the triviality of $\Pi_n(S^1)$ for $n>1$.  

Because it is useful to have mathematically-smooth maps, we would like to convert the Dirac strings of equation (\ref{dirac}) into a smooth map.  This we will do by converting some topological integrals into a form using Abelian gauge potentials \cite{oj,jp} and magnetic fields.  Since the field is literally the curl of the potential, it must be divergence-free, meaning that it has zero flux.  The zero-flux condition simply equates the flux from Dirac strings to the flux from a smooth monopole field, and thus preserves topological charge.  So we can express the topological charge via the smooth map rather than by accounting for  Dirac strings.

Refs \cite{oj,jp} show that the Chern-Simons number $N_{CS}$ for a static non-Abelian potential can be written in terms of an Abelian potential ${\mathcal{A}}_i$ and magnetic field $\vec{{\mathcal{B}}}=\vec{\nabla}\times \vec{{\mathcal{A}}}$.  In the present case the non-Abelian potential is pure-gauge: $A_i=U\partial_iU^{-1}$.   One finds:
\begin{equation}
\label{ojjp}
N_{CS}\equiv \frac{1}{24\pi^2}\int d^3x \epsilon_{ijk}Tr(U\partial_iU^{-1}U\partial_jU^{-1}U\partial_kU^{-1})=
\frac{1}{16\pi^2}\int d^3x\vec{{\mathcal{A}}}\cdot\vec{{\mathcal{B}}}
\end{equation}
where the Abelian potential and field are:
\begin{equation}
\label{abfp}
{\mathcal{A}}_j=iTr(\tau_3U\partial_jU^{-1});\;{\mathcal{B}}_j=-i\epsilon_{jkl}Tr(\tau_3U\partial_kU^{-1}U\partial_lU^{-1}).
\end{equation}
We have ignored any Dirac strings in writing the Abelian magnetic field.  This will not change the general topological integrals we write, except that it is not necessarily the case that $\vec{\nabla}\cdot\vec{{\mathcal{B}}}=0$.  This fact will lead to a cautionary note for the map $S^2\rightarrow S^2$ discussed next.

\subsection{The map $S^2\rightarrow S^2$}

The appearance of a $\tau_3$ projection in the Abelian quantities of (\ref{abfp}) suggests a close connection with the projection occurring in the monopole flux integral (\ref{charge}), in which the matrix $2Q_b$ is $\tau_3$.  The Abelian magnetic field of (\ref{abfp}) can be expressed in terms of a unit vector $\hat{n}$, defined as an $SU(2)$ rotation of $\tau_3$ by the gauge matrix $U$.  
In the case of  the standard pure-gauge nexus this gauge matrix becomes the matrix $V$ defined in  equation (\ref{veq}), leading to the $S^2\rightarrow S^2$ map:
\begin{equation}
\label{s2s2}
V\tau_3V^{-1}\equiv \hat{n}\cdot \vec{\tau};\;\hat{n}=\cos \phi\; \hat{z}+(1-\cos \phi )\cos \theta \; \hat{r} +\sin \phi \sin \theta \; \hat{\phi}.
\end{equation}
The index $I$ of this map is an integer, corresponding to the homotopy $\Pi_2(S^2)\simeq Z$, and is measured through the volume two-form and integral for  $I$.  The appropriate two-form is just the Abelian magnetic field $\vec{{\mathcal{B}}}$:
\begin{equation}
\label{omega}
{\mathcal{B}}_i=\frac{1}{2}\epsilon_{ijk}\epsilon_{abc}\hat{n}^a\partial_j\hat{n}^b\partial_k\hat{n}^c;\;I=\frac{1}{4\pi}\int d^2S_i {\mathcal{B}}_i.
\end{equation}
For future reference we note that the magnetic field lines of $\vec{{\mathcal{B}}}$, the solution of $\vec{dx}\times \vec{{\mathcal{B}}}=0$, are found from $d\hat{n}=0$ as one finds by writing out the field-line differential equations.  That is, these field lines are found as the intersection of the two surfaces defined by $\hat{n}=\hat{n}_0$, where $\hat{n}_0$ is any unit vector constant in space.  For the vector $\hat{n}$ of equation (\ref{s2s2}), which is independent of $r$, the equation  $\hat{n}=\hat{n}_0$ has straight lines emerging from the origin as  the solution.

Since ${\mathcal{B}}_i$ is closed (divergence-free) the index $I$ should be zero, strictly speaking.  That it is not zero in certain circumstances reflects the effect of Dirac string contributions which were ignored in the definition of ${\mathcal{B}}_i$.  (To speak in more legitimate mathematical terms, the second (co)homology group of $S^2$ is non-trivial, so closed two-forms need not be exact.)  Typically this effect shows up as a singular contribution to $\nabla_i{\mathcal{B}}_i $ in the form of a Dirac delta function at the origin of the two-sphere embedded in three-space. 

We now show that the index $I$ is the topological charge integral (\ref{cjeq}), but with the Dirac string contributions to $\vec{B}$ replaced by an equivalent amount of smooth flux.   Go back to the defining equation (\ref{abfp}) for ${\mathcal{A}}_i$ and this time save the Dirac-string contributions in $\vec{\nabla}\times \vec{{\mathcal{A}}}$; we call the result $\Omega_i$: 
\begin{equation}
\label{babel}
\Omega_i=\epsilon_{ijk}\partial_j{\mathcal{A}}_k=-i\epsilon_{ijk}Tr(\tau_3V\partial_jV^{-1}V\partial_kV^{-1})-2\pi\hat{z}\epsilon (z)\delta (x) \delta (y)\equiv {\mathcal{B}}_i+B_i(Ds).
\end{equation}
 The  Dirac-string field $B_i(Ds)$ is precisely the non-Abelian magnetic field $\vec{G}$, projected with $\tau_3$, given in equation (\ref{dirac}) which appears in the topological charge integral (\ref{charge}):
\begin{equation}
\label{bg}
B_i(Ds)=iTr(\tau_3G_i).
\end{equation}  
 (Note that the term ${\mathcal{B}}$ occurring in the Abelian field strength of (\ref{babel}) is cancelled in the non-Abelian field $G_i$ by the commutator term of the non-Abelian gauge potential.)
The divergence $\nabla_i\Omega_i$ is strictly zero, as one expects for a curl, and in this vanishing divergence the contribution of  ${\mathcal{B}}_i$ cancels that of $B_i(Ds)$.  It follows that the flux integrals are equal, so that the topological charge $Q$ in (\ref{cjeq}) is that of ${\mathcal{B}}_i$:
\begin{equation}
\label{equalf}
\int d^2S_i{\mathcal{B}}_i=\int d^2S_iB_i(Ds);\;Q=\frac{1}{4\pi i}\int d^2S_i{\mathcal{B}}_i.
\end{equation}

Explicitly, for the map using $V$ of equation (\ref{veq}) one finds 
\begin{equation}
\label{abela}
i{\mathcal{A}}_i=\frac{\hat{\phi}_i}{\rho}[\cos \theta+(\cos \phi -1)\sin^2 \theta ] -\sin \phi \sin \theta \frac{\hat{\theta}_i}{r};
\end{equation}
\begin{equation}
\label{abelb}
i{\mathcal{B}}_i=\frac{\hat{r}_i}{r^2}(\cos \phi -1)(1+2\cos \theta ).
\end{equation}
In these equations, only the first term on the right-hand side of the potential (equation (\ref{abela})) contributes to the flux or topological charge, which is unity.  This first term by itself describes a Wu-Yang monopole in a string gauge, with two half-strings pointing from the origin in either direction, just as occurs for a nexus. The only term in ${\mathcal{B}}_i$ which contributes is $-\frac{\hat{r}_i}{r^2}$, which corresponds to the radial field of the Wu-Yang monopole.

Of course, the physical nexus fields look nothing like the Abelian versions (\ref{abela}, \ref{abelb}), but we are assured that these smooth fields have the same topological charge as the nexus, or its pure-gauge version.

One must be careful when using Abelian magnetic fields which are not divergence-free, as we have done by using the field ${\mathcal{B}}_i$.  Consider the Chern-Simons integral (\ref{ojjp}).  Clearly, from (\ref{abela},\ref{abelb}) the Chern-Simons integrand identically vanishes, as appropriate for a map $S^2\rightarrow S^2$.  However, certain kinds of gauge transformations can change this zero by an integer amount.  These  gauge transformations arise by right-multiplying the gauge matrices $U$ or $V$ by $SU(2)$ rotations around the three-axis, which leave the unit vector $\hat{n}$ and field ${\mathcal{B}}_i$ unchanged, and change ${\mathcal{A}}_i$:
\begin{equation}
\label{csgauge}
V\rightarrow Ve^{i\tau_3\alpha (r)/2},\;{\mathcal{A}}_i\rightarrow {\mathcal{A}}_i +\partial_i\alpha (r).
\end{equation} 
We will require
\begin{equation}
\label{newgauge}
\alpha (r=\infty )=2\pi N,\;\alpha (r=0)=0
\end{equation}
which insures that the gauge transformation can be compactified at infinity ($R^3\rightarrow S^3$) and is non-singular at the origin.  A quick calculation shows that the Chern-Simons number of (\ref{ojjp}) changes by $N$.  This simply means that for maps of the type $S^2\rightarrow S^2$ the appropriate gauge-invariant index is the $\Pi_2(S^2)$ index of equation (\ref{omega}).

\subsection{The map $S^3\rightarrow S^2$ and the Hopf fibration}

Since the general map of the form (\ref{s2s2}) (that is, of an element of $SU(2)$ to a unit vector) is a map $S^3\rightarrow S^2$, one might have expected to find the homotopy of this map, which is $\Pi_3(S^2)\simeq Z$.  
Indeed, many authors have discussed this homotopy, corresponding to the Hopf fibration, in monopole/caloron studies \cite{ct2,ct3,oj,kvb,bhvw,pvb}.
Taubes \cite{ct1,ct2,ct3} has shown, using Morse theory, that there are special non-BPS solutions to the $SU(2)$ gauge-Higgs classical equations, which have the character of monopole-anti-monopole pairs, and identified the appropriate non-contractible loop in gauge-orbit space.  He  has explicitly exhibited \cite{ct2} a representative of the non-contractible loop, and interpreted it as a relative rotation by 2$\pi$ of the gauge orientation of a monopole with respect to the anti-monopole.  Van Baal \cite{pvb} has invoked the same arguments for the caloron.

There is a general reason to believe that the Hopf fibration plays a role in nexus-vortex topological charge, which is that nexus-vortex topological charge corresponds to a linking number, as does the index of the Hopf fibration.  Generally speaking, this linking number is that of a closed two-surface (vortex) and closed line (nexus world line) in d=4, but for certain special cases of vortices and nexuses this linking number is that of two curves in d=3.

We give the explicit linking-number formula for topological charge (which also furnishes a simple proof that Abelian intersection numbers cancel in pairs for closed two-surfaces in d=4).The pure-gauge part of the gauge potential for a center vortex characterized by the closed surface $S_V$ is, from equation (\ref{cv}),
\begin{equation}
\label{pot}
A_{V\mu}(x)=\left( \frac{-2\pi Q_c}{i}\right) \epsilon_{\mu\nu\alpha\beta}\partial_{\nu}\oint_{S_V} \frac{1}{2}d\sigma_{\alpha\beta}\Delta_0(x-z(\sigma ).
\end{equation}
We will consider the linkage of this surface with a nexus world line $\Gamma_N$, where this world line  separates center-vortex flux $Q_a$ on one side and $Q_b$ on the other, and show that the topological charge as given in equations (\ref{qno}, \ref{xno}) is given by the integral
\begin{equation}
\label{topolink}
Q=\frac{1}{2\pi i}\oint_{\Gamma_N}dx_{\mu}Tr[(Q_a-Q_b)A_{V\mu}(x)].
\end{equation}
This follows very simply by transforming the term in equation (\ref{topolink}) which involves the trace factor $TrQ_cQ_a$ with Stokes' theorem, using a surface $S_a$ whose boundary is $\partial X_a=\Gamma_N$, and the other trace factor $TrQ_cQ_b$ using a Stokes' theorem surface $S_b$ with boundary $\partial S_b=-\Gamma_N$.  These surfaces will be chosen to coincide with the surfaces of the center vortex on either side of the nexus world line (cf Fig. \ref{nexusfig1a}); the sum $S_a+S_b$ of the two surfaces has a single orientation.  Then using Stokes' theorem and the explicit form of $A_{\mu}$ in equation (\ref{pot}) $Q$ becomes:
\begin{equation}
\label{stokes1}
Q= Tr(Q_cQ_a)I(c,a)-Tr(Q_cQ_b)I(c,b)
\end{equation} 
where $I$ is the intersection number of the surfaces indicated, as given in equation (\ref{xno}). (Note, by the way, that this gives a simple proof that the intersection number of two closed surfaces of a single orientation vanishes, since $Q$ above vanishes when the individual trace factors are replaced by unity.)

Now each of the intersection numbers in equation (\ref{stokes1}), which involves the intersection of a closed surface $S_V$ and an open surface $S_{a,b}$, is the same linking number $L$, which can be written in the usual Gauss form by undoing Stokes' theorem:
\begin{equation}
\label{linkno}
L=\oint_{\Gamma_N}dx_{\mu}\oint_{S_V}\frac{1}{2}d\sigma_{\alpha\beta}\epsilon_{\mu\nu\alpha\beta}\partial_{\nu}\Delta_0(x(\Gamma_N)-x(S_V)).
\end{equation}   
It follows that the topological charge for this configuration is:
\begin{equation}
\label{q2}
Q=LTr[Q_c(Q_a-Q_b)]=L(\delta_{ac}-\delta_{ab}).
\end{equation}
This is an integer, which can be zero either if $a=b$ (no nexus) or if the vortex linked to the nexus has a different $Q$ from either part of the nexus.

    Since we are dealing with topological quantities, it is no loss of generality in the link formula (\ref{linkno}) to take the vortex surface $S_V$ to be a cylinder which is the product $\Gamma_V\times x_4$, with the closed contour $\Gamma_V$ and the surface of the center vortex carrying the nexus both embedded in the $x_1,x_2,x_3$ volume.  In that case the integral over $x_4$ in (\ref{linkno}) can be done, and the link number $L$ becomes a d=3 link number of two contours:
\begin{equation}
\label{link3}
L=\oint_{\Gamma_V}dx_i\oint_{\Gamma_N}dy_j\epsilon_{ijk}\partial_k\Delta^{(3)}_0(\vec{x}-\vec{y})
\end{equation}
where $\Delta^{(3)}_0$, the integral over $x_4$ of $\Delta_0$, is the d=3 inverse of the Laplacian.  Furthermore, as is well-known this link number is an Abelian Chern-Simons form:
\begin{equation}
\label{cherns}
L=\int d^3x \vec{A}_V\cdot \vec{B}_N
\end{equation}
with
\begin{equation}
\label{cherns1}
\vec{A}_V(\vec{x})=\vec{\nabla}\times \oint_{\Gamma_V} d\vec{z}\Delta^{(3)}_0(\vec{x}-\vec{z})
\end{equation}
 for and
\begin{equation}
\label{cherns2}
\vec{B}_N(\vec{x})=\oint_{\Gamma_N} d\vec{z}\delta (\vec{x}-\vec{z}).
\end{equation}
Note that this Chern-Simons form is not the canonical one arising from the fact that the topological charge density $\sim G\tilde{G}$ can be written as a total divergence; for that canonical form, the topological charge emerges as the difference of two Chern-Simons integrals.  Nor is the quantity denoted $\vec{B}_N(\vec{x})$ actually, as the notation might indicate, the magnetic field of the nexus.  It is, however, of the form $\vec{\nabla}\times \vec{A}_N$, with $\vec{A}_N$ in the same form as in equation (\ref{cherns1}) but with $N$ replacing $V$.  Since (in the simple cases considered here) neither the vortex nor the nexus has any self-linkage one can write the link number as:
\begin{equation}
\label{totallink}
L=\frac{1}{2}\int d^3x \vec{A}\cdot \vec{B};\;\vec{A}=\vec{A}_V+\vec{A}_N;\;\vec{B}=\vec{\nabla}\times \vec{A}.
\end{equation}

The Chern-Simons integral $L$ is a Hopf invariant, and so presumably it would be possible to find an explicit Hopf fibration which contains, in its family of linked curves, the curves of the nexus and the center vortex.  We will not pursue that issue here.   

\section{A new nexus which carries Chern-Simons number}
\label{anewnexus}

So far we have considered a standard nexus which has topological charge by virtue of a linkage between the nexus worldline and a center-vortex surface.  No Chern-Simons number was associated to this topological charge.  It is also possible to find a different d=3 nexus which carries Chern-Simons number quite independent of any vortex linkages, and that is our present subject.  This new nexus depends on all three space variables $r,\theta ,\phi$ and so corresponds to a map $S^3\rightarrow S^2$.

Consider the pure-gauge nexus described by the matrix $R$:
\begin{equation}
\label{newnexus}
A_i=R\partial_iR^{-1};\;R=e^{i\phi \vec{\tau}\cdot \hat{e}};\;\hat{e}=(\sin \theta \cos \beta (r),\sin \theta \sin \beta (r), \cos \theta ).
\end{equation}
We will require that the angle $\beta$ runs from 0 to  $2\pi$ as $r$ runs from zero to infinity.
The gauge potential of (\ref{newnexus}) yields the Dirac-string field strengths (equation (\ref{dirac})) of the static $r$-independent pure-gauge nexus we have been studying up to now.  The matrix $R$ yields a transformation $S^3\rightarrow S^2$, which defines a unit vector $\hat{m}$: 
\begin{equation}
\label{hopf}
R\tau_3R^{-1}=\vec{\tau}\cdot \hat{m}.
\end{equation}  
The explicit form of $\hat{m}$ is:
\begin{equation}
\label{meq}
\hat{m}=\hat{z}\cos \phi +\hat{\beta}\sin \theta \sin \phi +\hat{e}(1-\cos \phi )\cos \theta ;\;\hat{\beta}=(-\sin \beta , \cos \beta ,0).
\end{equation}   
 As before, the family of linked (twisted, writhing) curves of field lines is given by $\hat{m}=\hat{m}_0$ as the unit vector $\hat{m}_0$, constant in space, is varied.
The explicit form of the Abelian potential and field of equation (\ref{abfp}) are:
\begin{equation}
\label{nabp}
{\mathcal{A}}_i=\cos \theta \frac{\hat{\phi}_i}{\rho}+(\cos \phi -1)\sin^2\theta \beta '\hat{r}_i -\sin \phi \sin \theta \frac{\hat{\theta}_i}{r};
\end{equation}
\begin{equation}
\label{nabf}
{\mathcal{B}}_i=(\cos \phi -1)\frac{\hat{r}_i}{r^2}-\beta '\sin \phi \sin \theta \frac{\hat{\theta}_i}{r}-2\beta '\sin \theta \cos \theta (\cos \phi -1)\frac{\hat{\phi}_i}{r}
\end{equation}
From equation (\ref{ojjp})  the Chern-Simons number of the gauge potential based on $R$ is unity:
\begin{equation}
\label{rcs}
N_{CS}=\frac{1}{8\pi^2}\int d\phi (1-\cos \phi )\int \sin \theta d\theta \int_0^{\infty}dr \beta '(r) = \frac{1}{2\pi}[\beta (\infty )-\beta (0)]=1.
\end{equation}
Note that the field lines in (\ref{nabf}) twist on their way to infinity (that is, to the anti-nexus at infinity), which is the source of the Chern-Simons number.  Two twisted field lines are, in fact, linked.

As a purely static configuration this new nexus by itself carries no topological charge in d=4, since this charge is the difference of two equal Chern-Simons numbers.  But it does carry d=4 topological charge when linked to a center vortex, just as the previously-considered nexus does, as one sees by calculating the $S^2\rightarrow S^2$ index of equation (\ref{omega}).  Moreover, one can extend the static new nexus to a dynamical configuration which does carry integral charge, even in the absence of a vortex.  This requires extending $\beta (r)$ to a function $\beta (r,t)$, where $t$ is the Euclidean time variable.  One possible choice is:
\begin{equation}
\label{newbeta}
\beta = 2 \tan^{-1}(r/t)
\end{equation}    
which approaches $2\pi$ at $t\rightarrow -\infty$ and 0 at $t\rightarrow +\infty$, thus yielding unity for the difference of the two Chern-Simons numbers given by
equation (\ref{rcs}).  Presumably this $SU(2)$ topological charge can be split into two fractional charges of 1/2 each, by extending the earlier work of \cite{ct} which shows how to split a conventional $SU(2)$ instanton into two such fractions.  Of course, quite aside from any d=4 considerations it is interesting to have d=3 configurations which play the roles both of center vortex and of carrier of Chern-Simons number. Earler work \cite{co95,co96,cy,c98} discussed the dynamics of reconnection processes involving linkage, twisting, and writhing of vortices leading to Chern-Simons number and topological charge; these are similar to what is happening with the dyamical new nexus.  One should expect to find static but unstable configurations in d=3 with Chern-Simons number of half an integer, corresponding to sphalerons, as discussed in the above-cited works.

\section{The $SU(N)$ case:  Complete fractionalization}
\label{frac}

First we discuss the generalization of the $SU(2)$ nexus of equation (\ref{trial}) to general $N$.  Then we show how to decompose this two-flux-tube nexus into $N$ tubes.

The generalization to any $N$ of equation (\ref{trial}) is straightforward; the $N=3$ case has been given explicitly.(There is a misprint in equation (13) of \cite{co00}; $\lambda_3$ in the exponent on the right-hand side should be replaced by $\lambda_1$ or $\lambda_2$ \cite{co00}.)  The fundamental nexus of $SU(N)$ has a flux along the negative $z$ axis corresponding to $Q_a$, and a flux along the positive $z$ axis corresponding to $Q_b$.    For any values of $a,b$ there is a 2$\times$2 subspace in the matrix space of the generators $\lambda_J$; this subspace has entries in the intersections of the $a$ and $b$ rows and columns.  Without loss of generality we can display the generators in a basis where $a=1,\;b=2$.  Introduce the Pauli matrices $\tau_j$ living in this (1,2) subspace, with zeroes in all other positions.  We write $Q_{1,2}$ as:
\begin{equation}
\label{2q}
Q_1=-\frac{1}{2}\tau_3+R(12);\;Q_2=+\frac{1}{2}\tau_3+R(12); R(12)=\rm{diag}(-\frac{1}{2}+\frac{1}{N},-\frac{1}{2}+\frac{1}{N},\frac{1}{N},\dots ).
\end{equation}
Note that $R(12)$ commutes with all the $\tau_j$ and is traceless; it is a linear combination of the $Q_a$ with $a\neq 1,2$.
Then the gauge $V$ for the $SU(2)$ case of equation (\ref{veq}) is to be replaced by 
\begin{equation}
\label{veq2}
V=e^{(i\phi \tau\cdot\hat{x}/2)}e^{i\phi R(12)}
\end{equation}
which has the correct holonomy $\exp (2\pi i/N)$.
This is just a combination of an $SU(2)$ nexus with a special vortex, described by $R(12)$.  This $R(12)$ vortex is not  a center vortex, but two of them taken together is, corresponding to center element $\exp (4\pi i/N)$.  (It is not a center vortex because $\exp (2\pi i R(12))$ is the product of a center vortex with the matrix diag(-1,-1,1,1,...).  The latter matrix does not commute with group generators $\lambda_{ij}$ where one of the indices $i,j$ is not in the (1,2) subspace but the other is.)
Using the decomposition of the $Q_a$ given in equation (\ref{2q}) and the corresponding $m=\infty$ description of the nexus in equation (\ref{dirac2}), one sees that the special vortex described by the flux matrix $R(12)$ drops out of the trace, leaving only the usual $SU(2)$ Pauli matrices.
Then the Dirac-string field strength of equation (\ref{dirac}) can be replaced by:
\begin{equation}
\label{dirac2}
i\vec{G}=- 2\pi \hat{z}\delta (x) \delta (y)[Q_2\theta (z)+Q_1\theta (-z)].
\end{equation}
The (of course, expected) result is that the $SU(2)$ formulas continue to hold for each of the $SU(2)$s imbedded in $SU(N)$.

\begin{figure}[ht]
\includegraphics{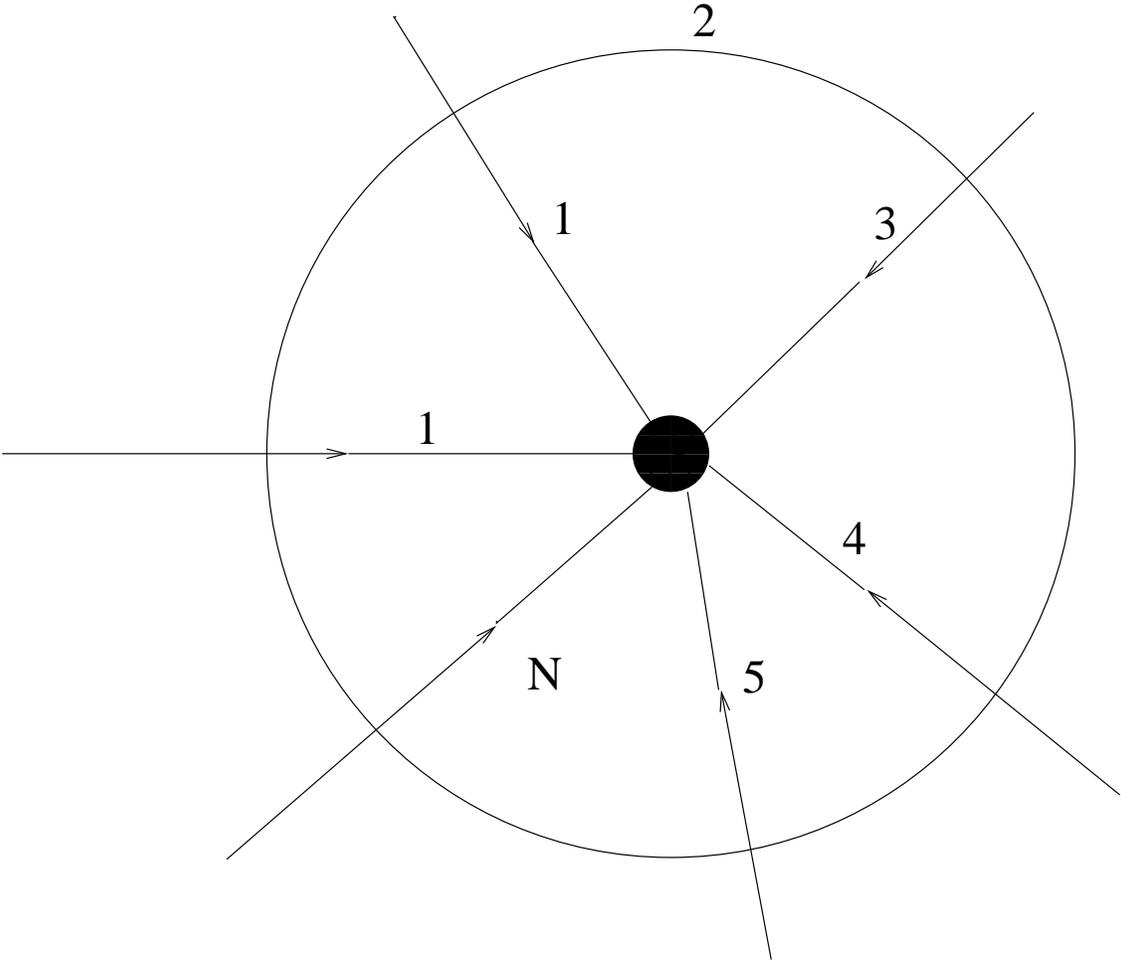}
\vskip.2cm
\caption[]{
\label{nexusfig2}
\small The flux tube labeled 2 in Fig. \ref{nexusfig1}
is split into $N-1$ oppositely-directed flux tubes.}
\end{figure}

The obvious question is whether one can further decompose these two quanta, ultimately ending with $N$ quanta each of topological charge $1/N$.  These could be called quarks of topological charge, because they are the smallest unit of topological charge and because they are confined topologically in such a way that the total charge is necessarily integral.  In fact, it is easy to find these smallest-charge constituents.  In Fig. \ref{nexusfig1} there is one nexus flux tube carrying the flux matrix $Q_2$.  Because the sum of all $Q_a$ is zero, we can write
\begin{equation}
\label{qsum}
Q_2=-Q_1-Q_3-\dots -Q_N
\end{equation}
which corresponds to the physical process of splitting the flux tube labeled 2 in Fig. \ref{nexusfig1} into $N-1$ tubes, with flux going in the opposite direction as specified by the minus signs on the right-hand side of (\ref{qsum}).  This is illustrated in Fig. \ref{nexusfig2}.

Of course, there is unit topological charge in this configuration but now there are, as desired, $N$ localized units of topological charge $1/N$ each.  They are localized to the points of intersection of the plain vortex (labeled 2 in Fig. \ref{nexusfig2}) and the flux tubes emerging from the nexus.

It might be objected that the configuration of Fig. \ref{nexusfig2} is not probable because it has $N$ flux tubes each of which has action equal to that of either flux tube of Fig. \ref{nexusfig1}, so that the $N$-quantum configuration suffers an action penalty.  But again we presume that the entropy of each flux tube exceeds its action, so that in fact the $N$-quantum configuration is preferred.  And either of the two configurations we have considered will be favored over a simple classical instanton, which does not have the geometric entropy of flux tubes and vortex surfaces.

\section{Summary and conclusions}
\label{conc}

We have shown that even without  adjoint scalar fields of various types which are invoked in the traditional viewpoint, similar or identical mathematical constructions appear both in the traditional viewpoint and in the center vortex-nexus viewpoint.  But the interpretation of these formulas is different in the two cases.  

The 2-surface at infinity introduced to measure monopole flux (magnetic charge) in the traditional viewpoint, exhibited in equation (\ref{cjeq}) is just a mathematical surface used for measuring magnetic flux.  But the surface of integration in equation (\ref{charge}) is physical; it is, in fact, a thick surface corresponding to a center vortex.  Topological charge is located not at the center of a BPS monopole but at the points where the nexus flux overlaps with the plain center vortex surface.  This is clear from the interpretation of the standard form of topological charge (equation (\ref{q}) as an intersection number, given by equation (\ref{xno}).  So the {\em integral}
topological charge measured by the integral (\ref{charge}) is composed of {\em localized and non-integral quanta}.   In the $SU(2)$ case these have topological charges $\pm$ 1/2.  The fundamental nexus analyzed in \cite{co00} for general $N$ also has two quanta, one of charge 1-$1/N$ and the other of charge $1/N$.

We showed that the same topological charge could be explicitly expressed as a linkage number of a center vortex two-surface with a nexus world line, and that in a special (static) case this linkage appeared as the linkage of two closed curves, which we interpreted in terms of the linkages occurring in the Hopf fibration.

In section \ref{anewnexus} we found a new nexus, which is, as for the standard nexus, a world line on a center-vortex sheet dividing regions of differently-directed field strength.  This new nexus world line can exhibit unit topological charge by linking with a center vortex.  It can also be interpreted as a static d=3 configuration carrying integral Chern-Simons number all by itself, with no reference to vortices.  It can be extended to a dynamic (time-dependent) configuration which carries unit topological charge in d=4 with no connection to vortices.  Further investigation of this dynamic nexus is  underway.

In Section \ref{frac} we analyzed the topological charge problem for general $N$, and showed how to decompose further into $N$ quanta each of topological charge $1/N$.  The same fractionalization occurs in the caloron, but the caloron appears to be a different sort of construction from a center vortex-nexus combination.  By virtue of standard QCD entropy arguments, the fractional quanta of the present paper are virtually independent of one another, that is, there are no short-range forces between them.  The only connection between them is by highly-sinuous lines and surfaces, lines which are very long in terms of the shortest distance between the two quanta.  This independence shows up in the topological susceptibility, which measures the mean-square topological charge \cite{co00}.  And just as the three quarks in a proton have total (electromagnetic) charge 1, their mean-square charge of 2/9 reveals the fractional nature of quark charge.  This is, of course, the kind of thing needed for consistency with the Witten-Veneziano sum rule.

It will be important to analyze the nature of the fermionic zero modes associated with this kind of topological charge.   A first step was taken in this direction \cite{ct} for $SU(2)$ topological charge units of charge 1/2 each, joined by a sphaleron-like world line.  There is no known generalization of the work of \cite{ct} to other gauge groups, nor is the connection between this earlier fractional topological charge and center vortex-nexus topological charge known.  This is also an interesting topic for future investigation.

There are, as shown above, numerous similarities of functional form between integrals used to express topological charge in the traditional viewpoint and in the center vortex-nexus viewpoint.  It is true that these have different physical interpretations in the two viewpoints, because one has adjoint scalars and the other does not.  Nevertheless, one may be encouraged to think that it will be possible to find further analogs between various points of view, such as the center vortex-nexus picture and the dual superconductor picture.  This is possible, but it will require considerable more investigation.  For example, in the semiclassical dual superconductor picture, there are $N-1$ massless gauge bosons, corresponding to the Cartan subalgebra.  It is, however, likely that quantum effects give all gauge bosons a finite mass, as has been argued \cite{c99} for the d=3 Georgi-Glashow model in the limit where the scalar VEV vanishes.

Finally, we believe that an important application of the work of Faddeev and Niemi
\cite{fn}, who introduce a decomposition of Yang-Mills potentials involving (among other things) unit vectors such as our vector $\hat{n}$ of equation (\ref{s2s2}) or $\hat{m}$ of equation (\ref{meq}), is to characterize the time-dependent dynamics of nexuses in interaction with other elements of the QCD vacuum.  (An example is the time dependence of the new nexus discussed in section \ref{anewnexus}.)  Work in this direction is in progress.

\end{document}